# Metrological support of acoustic measuring installations mid-frequency devices

A.N. Grekov, N.A. Grekov, E.N. Sychev

The article discusses methods for measuring the speed of sound, scattering, attenuation and absorption of sound in liquids, on the basis of which the structural diagrams of modern devices have been developed. In real conditions, acoustic measurement schemes are not ideal and depending on specific structure may give different results. The technical and metrological characteristics of the measuring channels of modern acoustic devices are presented. Recommendations are given for the use of GOST in acoustic measurements and it is stated that the metrological characteristics of newly created measuring instruments operating in in situ conditions are at the level of State primary standards.

Introduction. Great variability physico-chemical and biological characteristics of inland water bodies, shelf zones of seas and oceans puts the task of operational monitoring of dynamics studied processes and improvement of environmental monitoring measurement systems.

Measured basic acoustics parameters in the liquid under study make it possible, for example, to determine complex hydrophysical parameters sea water of the oceans and seas, use them to control various impurities not only in sea water, but also in river and waste water and to study the physical properties of water, mixtures of liquids and clarify empirical expressions in determining thermodynamic quantities and molecular properties liquids.

It is known that the advantage of acoustic methods for measuring the speed of sound propagation, attenuation, scattering and absorption of ultrasound is that the transparency of the liquid does not matter, unlike the optical method, where only optically transparent liquids are studied. In the presented work we consider the mid-frequency acoustic range. At the same time, when considering physical processes, we partially move to the beginning of the high-frequency range in the understanding that that range boundaries are arbitrary concept. Measuring acoustic systems operating in situ are constantly being improved and their development together with metrological support are priority areas marine instrument making. Speed of propagation of ultrasound. Let's briefly look at a few methods for measuring the speed of sound in liquids on which the structures of measuring instruments are based. One method for measuring c may be to determine the wavelength $\alpha = \frac{2\pi}{k}$ at frequency $v = \frac{\omega}{2\pi}$

Another method is to obtain the speed of sound from the time interval t, which is measured at a short pulse travels a distance x. More details on this question considered in [1], where it is noted that that the group velocity in the medium is equivalent to the phase velocity only when no dispersion. For liquids, in which there is relaxation, these the speeds will not be equal. In some cases of differences between phase and group velocities exceed the accuracy of the velocity meter used, so they must be included in quantitative group velocities speed. Good measurement performance sound speed is given by the continuous method waves. Based on this method, the speed of sound can be determined in two ways. In the first of them, the measuring cell is located between a pair of transducers or between a transducer and a reflector, which is a resonator [2]. In such meters, the gear ratio is determined function of the environment, and from it they calculate sound speed. Such devices are good have proven themselves in laboratory settings when determining speed sound in small volumes of matter. The achieved accuracy of resonator methods is 1·10-7. Slow performance and complexity makes meters based on this method unsuitable for in situ use. In the second method to determine speed of sound uses the interaction of continuous acoustic waves with optical waves [3]. The measurement accuracy achieved by these methods is 1·10-6 [4]. Disadvantages These methods are difficult equipment and inapplicability of the method for opaque liquids. The pulse method of measuring the speed of sound has become widely developed. This method, in turn, is divided into two: narrowband and broadband pulse methods. In the first case, as an emitted signal an amplitude-modulated harmonic signal is used. Strictly speaking, its frequency response is not will appear as a sharp peak at a fixed frequency.

With a carrier frequency of 10 MHz, the pulse duration for this method ranges from one to tens of microseconds [5]. This method has found wide application in systems measuring the speed of sound in the test area liquid by comparison with a reference speed of sound [6].

In broadband pulse methods, the signal used is pulses with duration less than 1 μs at carrier frequencies of 1–10 MHz. In such meters, speed determination sound lies in precise measurement time Δt required for the signal to travel the distance Δx.

Several methods have been developed for measuring time Δt for pulse methods. These are the pulse-cyclic method, the correlation method and the direct measurement of propagation time. In systems with a pulse-cyclic method, the accuracy of measuring the speed of sound has been achieved 1·10-7 [7].

Essentially, in such meters for obtaining information about propagation time requires multiple repetition of the procedure for transmitting and receiving an acoustic signal, which reduces the performance of the system. The best the correlation method is fast; the measurement accuracy achieved by this method is 1·10-6 [8]. Measurement methods considered sound speeds, with the exception of pulsed ones, have not found application in real life measuring instruments produced by the world's leading manufacturers in situ due to the shortcomings noted higher. Therefore, in modern devices pulse method is widely used direct time-of-flight measurement acoustic signal. This method has good performance, satisfactory accuracy and simplicity. Implementation. For low-viscosity liquids, which include sea water, the upper frequency range of acoustic waves, in which is conducting research on the speed of sound c, currently reaches 1·1010 Hz. Technically, to obtain Thin piezoelectric and piezosemiconductor films are used at such frequencies [9]. In real conditions use for water research frequencies above 1·109 Hz is difficult, since measurements must be carried out at distances of hundreds of microns and pa surf the waves with great intensity. Produced speed meters sound, based on the pulse method in conjunction with measuring temperature and pressure channels, widely used in profilers (SVP profilers) for research aquatic environment in situ. Measuring channels equipped with SVP profilographs and the achieved accuracy sound speed temperature measurements and pressures through these channels reflect the achieved technical level of the developers of these devices. Let's give measurement range and accuracy achieved by Valeport Limited in their SVP devices [10].

Technical and metrological characteristics of measuring channels speed of sound, temperature and pressure for acoustic signal frequency 2.4 MHz are given in table. 1.

Table 1. Technical and metrological characteristics of the measuring channels of SVP profilographs

| Sensor | Permission | Range | Calibration error | Stability | Final Accuracy |
|---|---|---|---|---|---|
| 25 mm | 0.001 m/s | 1400 – 1600 m/s | ±0.01 m/s | ±0.085 m/s | ±0.095 m/s |
| 50 mm | 0.001 m/s | 1400 – 1600 m/s | ±0.006 m/s | ±0.054 m/s | ±0,06 m/s |
| 100 mm | 0.001 m/s | 1400 – 1600 m/s | ±0.003 m/s | ±0,027 м/с | ±0,03 m/s |
| Pressure | 0.001 % | 0 to 100, 500, 1000 or 6000 dBar | | | ±0,05% (от 10 до 40°C) |
| Temper | 0.001 C$^0$ | -5 ÷ +35°C | | | ±0,01°C |

Temporary resolution is 1/100 ns, which is equivalent to a speed of sound of approximately 0.5 mm/s at the measuring base 25 mm, or 0.125 mm/s base 100 mm. According to the manufacturers, practically achieved resolution 1 mm/s. SVP profilers with additional measuring channels can be used for various sounding and buoy systems that designed for research in oceans and seas.

In Russia, JSC Concern Okeanpribor has developed and is manufacturing a sound speed profiler called MG-543EM. This profiler is used for work at depths of up to 350 m. Sound speed measuring channel the MG-543EM device is significantly inferior in accuracy to foreign analogues, the error of which is Specialists from the Federal State Budgetary Institution "Institute natural-technical systems", having extensive experience in marine development devices, created a sound speed profiler called AIS-1 [11]. Technical characteristics of the profiler (probe) IZZ-1 are presented in table 2. Primary transducer of the sound speed measuring channel of the device AES-1 has its own design features, which are: that it has two reflectors. One of the reflectors has a translucent design for acoustic signal, which eliminates the influence of hydrostatic pressure on the measuring base and does not require additional pressure calibration. Having a high-quality measurement base, which is formed due to highly stable sital rods or carbon composite material, good metrological characteristics were obtained, including time stability and accuracy speed of sound. Using various materials of the profilograph's protective housing, as well as using various modifications of pressure sensors installed in the device, it is possible to carry out measurements at depths of up to 8000 m.

Table 2. Metrological characteristics of the ISZ-1 profilograph

| Measured parameters | Measurement range | Random error | Error |
|---|---|---|---|
| Speed of sound propagation in liquid, m/s | 1400 ÷ 1700 | 0,001 | ± 0,02 |
| Water temperature, °C | -2 ÷ +35 | 0,001 | ± 0,01 |
| Hydrostatic pressure, dbar | 0 ÷ 8000 | 0,8 | ± 8 |

Operation of speed meters sound is inextricably linked with metrological support, and these issues were always relevant, and their solution based on modern achievements in Science and Technology. Metrological security sound speed meters in liquids, working not only in laboratory conditions, but also in situ, not fully meets modern requirements, so

How are newly created acoustic measuring instruments positioned at State primary level standard for the unit of sound speed in liquid media. For metrological support devices developed by SVP in Russia there is a State primary standard that reproduces the unit speed of sound propagation in liquid media with root mean square deviation of measurement results, not exceeding 0.005 m/s with 15 independent measurements and non-excluded systematic error not exceeding 0.04 m/s with confidence probability P = 0.99.

The standard uncertainty of the measurement result by the standard does not exceed: for type A – 0.005 m/s; by type B – 0.02 m/s; total – 0.02 m/s.

Expanded uncertainty of the measurement result with a standard at k = 2, U= 0.04 m/s.

The secondary standard reproduces unit of speed of sound propagation in sea water with a standard deviation of measurement results not exceeding 0.05 m/s at 15 independent measurements and non-excluded systematic error, not exceeding 0.08 m/s with a confidence probability of P = 0.99. Secondary standard can be used to calibrate profilers in exceptional cases mainly for marine conditions. Presented in GOST R 8.870- 2014 working standards of the 1st, 2nd category, and also the working measuring instruments go beyond the error limits of the SVP measuring instruments created and used in situ. In order to save money on metrological support for the ISZ-1 device in our practice, to calibrate the sound speed measuring channel, we used a well-known method, which is based on the totality high-precision experimental data for atmospheric pressure [9, 12– 15]. The validity of this approach is due to the fact that the design, or rather

The measuring base of the primary sound speed transducer IZZ-1, as noted above, does not depend on pressure. The calibration of AES-1 was carried out in thermostat with normal deaerated, distilled water, where the device was completely immersed. Day off The N code of the device was recorded at various temperature levels within from 4 to 28°C [16]. Then using the output value code N for the corresponding values temperatures, according to experimental tabular data on the dependence of the speed of sound on temperature were determined, using the least squares method, the coefficients of the regression curve, the graph of which is presented in rice. 1, a. The "c" axis shows the speed of sound depending on the temperature, and along the "N" axis – the output value of the code. In Fig. 1, b shows the discrepancy values (Ns) of the regression curve, obtained during calibration, and tabulated speed values sound "s" for distilled water. Standard deviation for regression curve is N0.01 m/s at the speed of sound.

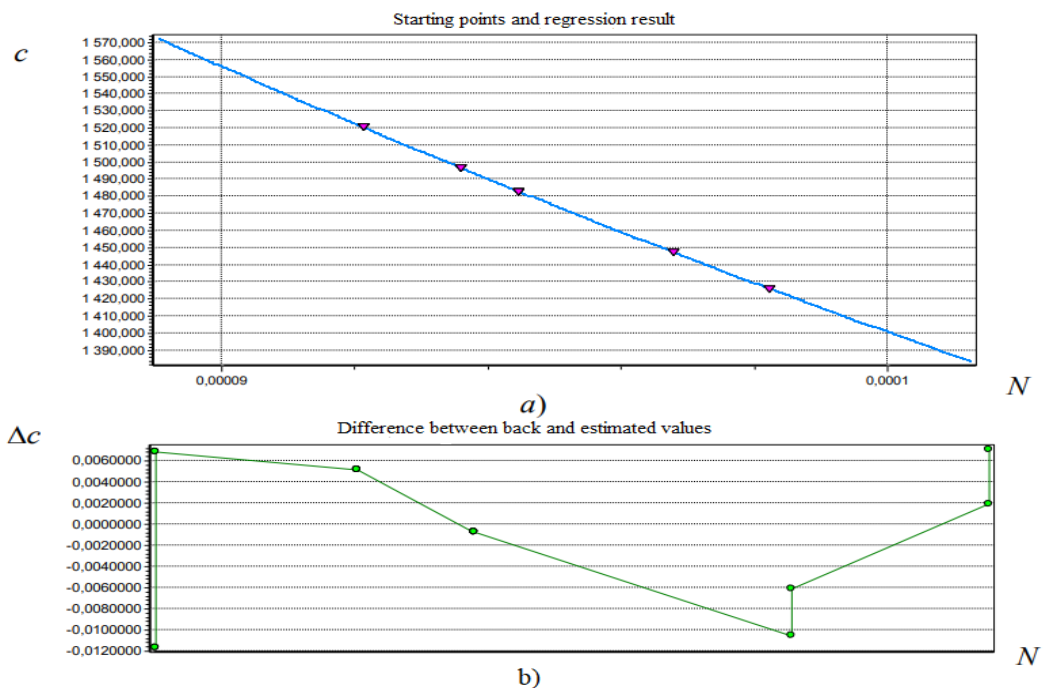

a) Speed of sound (s), m/s, and code value N; b) the magnitude of the discrepancy between the obtained values speed of sound (s) from experimental table values (Δs), m/s

Fig. 1. Results of calibration of satellite-1

Analysis of the latest publications in the field of ensuring uniformity of measurements the speed of sound in the aquatic environment showed a large discrepancy between the errors obtained at VNIIFTRI on State primary standard units of sound speed in liquid media and data presented in the passports of devices from foreign manufacturers (Table 3) [17].

Table 3. Accuracy characteristics of sound speed meters [17]

| Manufacturer and name devices | Devices Errors according to speed of sound, specified in the ED, m/s | Limits of permissible speed errors sound defined after testing for type assertions |
|---|---|---|
| 1. SVP-20. SVP-25 RESON, DENMARK | 0.25 | 0.6 m/s |
| 2. SVP-70, SVP-75 RESON, DENMARK | (0.05…0.25) | 0.25 m/s |
| 3. Midas SVP Valeport, UK | 0.02 | not studied |
| 4. Midas SVX-2 Valeport, UK | 0.02 | not studied |
| 5. miniSVP Valeport, UK | 0.02 | 0.25 m/s |
| 6. miniSVS Valeport, UK | 0.017 | not studied |
| 7. Digibar Pro ODOM, USA | 0.1 | not tested |

Ultrasound scattering. Measurement Ultrasound scattering even in laboratory conditions is a difficult task for a number of reasons. First of all, don't It is always sufficient to simply distinguish between absorption and scattering, sometimes a local case of acoustic attenuation can be treated as as a limiting case of scattering. Secondly, in real conditions any measurement schemes are not ideal and depending on the specific structures can produce different results.

Acoustic devices, about which discussed in this article are designed to work in situ and it is natural that number of measured scatterers in there should be a lot of liquid and they randomly distributed in the study environment. In this case, in the Born approximation, the total scattering cross section is determined by the expression

$$\mu_s = \frac{1}{V}\{k_i^2[\beta(k) + q(k)cos\varphi]\}^2 \, d\theta sinxdx \qquad 1$$

Где

$$cos\varphi = {k_i \cdot k_s}\big/{|k_i|^2}$$

Where $\varphi$ is the angle between the directions of the incident and scattered waves; $k_i$ – wave vector of the incident wave; $k_s$ – wave vector of the scattered wave; $\beta(k)$ and $q(k)$ – density and compressibility moduli. Angles and are the spherical coordinates of the wave vector of the scattered waves $k_s$ relatively; $k_i$ V – valuescattering volume

In environments where scatterers are randomly distributed and the object in question contains a sufficiently large number of them, the scattered power will be proportional to this volume, and the scattering is incoherent. In our case, the magnitude of scattering can b characterized using the scattering cross section per unit volume. This quantity characterizes the scattering coefficient and has the dimension m-1 The differential coefficient is determined in a similar way scattering A special case of a differential section is the inverse section scattering oil products com in wastewater (based on PNDF 14.1:2.116);

– Methodology for determining the content of impurities (suspended substances) in

wastewater (based on PNDF 14.1:2.110). The methods are certified by the Federal State Unitary Enterprise UNIIM Rostekhregulirovaniya.

From Western meters you can ote the MONITEK ULTRASONIC™ OIL IN WATER & TSS device, designed and manufactured in Massachusetts, USA. For measurements use also acoustic inverse method scattering at frequencies of 5 and 15 MHz. The device can detect particles bubbles and suspended droplets in various flowing liquids.

Metrological characteristics of the device:

– basic reduced error ± 4%;
– sensitivity ± 2%;
– response time 1 s;
–measuring range / parameters from

0 to 10 mg/l to 3% / suspended solids, oil in water. The device can read up to 16 various samples and using this information, build a lookup table or perform a polynomial fit of powers 1 to 4. User can plot the resulting curve graphically to visually verify its consistency with samples. Samples can be edited or disabled. Methodology calibrations developed by the device manufacturer based on the 0–1000 ppm model for frequencies 5 MHz and 15 MHz using diatomaceous earth. Typically 5 calibration points are used, which are determined by stirring 2.0 g diatomaceous earth in two liters of deionized water, which is 100% of the range (1000 ppm) of the instrument scale. Sample constantly stir. Then diluted in the following concentrations: 75, 50 and 25% and poured into containers, where the acoustic transducer is sequentially lowered and calibrated according to reference points. Regarding these calibration points are compared with the measured solution. In these instruments lack measuring sound speed and attenuation channels, which are limits the range of their application. In IPTS based on a speed meter sound system IZZ-1, a prototype of the device was developed with attenuation measurement and an additional acoustic measuring channel, the ultrasonic receiver of which located at an angle to the incident beam, which allows you to simultaneously measure backscattering and forward scattering and improve the accuracy of the measured parameters. At present, a verification scheme with standards for determining the acoustic scattering coefficient in liquids has not been created. And there are objective reasons for this, since the complexity the determination of this coefficient is related to the diversity of the liquids themselves, the shape and composition of the diffusers and the variety of different hardware implementation, which was briefly discussed higher. Therefore, in the current conditions, it is possible to use the developments presented in GOST R IEC 62127-1-2009 GSI [18], GOST R IEC 61161-2019 [19], and GOST R IEC 61391-2-2012 GSI [20].

These GOSTs show that in laboratories involved in backscattering coefficient measurements, level of accuracy can be achieved required to create calibrated reference phantoms [21–23]. In practice, reference samples contain scattering targets with more simple frequency dependence and this allowed by current standards [24, 25]. Information about the dissipation coefficient, driven not at 1 MHz, but at 3 MHz, correspond to small and acceptable variations in echo signal levels obtained for samples with weak frequency dependence of the coefficient scattering, provided that the attenuation in This sample complies with the requirements of this standard. Evaluation of local dynamic range can be performed when scanning a phantom containing spherical or cylindrical objects you, backscatter coefficients which differ from the main one sample to a known value [26]. IN

In some phantoms, the backscatter coefficients of these objects vary by up to 24 dB (by 15 dB more than the main sample, and by 9 dB less).

Phantoms suitable for measurement are based on aqueous gels with microscopic inhomogeneities, evenly distributed throughout the volume so as to provide the required level of attenuation [24, 25, 27–30]. Such phantoms must also contain solid particles such as glass beads (with a through hole) with a diameter of 40 microns, which give backscatter signals of controlled amplitude [30, 31]. Many manufacturers (for example, ATS Labs, Bridgeport, CT, USA (www.atslabs.com), CIPS, Norfolk, VA, USA (www.cirsinc.com), Gammex/RMI, Middleton, WI, USA (www.qammex.com)) produce these products and their phantoms, meeting these requirements. Ultrasound attenuation. It is known that the attenuation is caused by a number of processes, such as absorption, scattering, reflection, refraction and wavefront divergence. Many processes affecting attenuation depend on structural features of primary converters and the frequency range in which they operate. Considering what we're looking at metrological support for acoustic measuring instruments operating in the mid-frequency range with measuring bases not exceeding tens of centimeters and with a practically flat reflector with a surface reflection coefficient close to 0.93, and the dimensions of which are significantly greater than the width of the ultrasonic beam at the location of the reflector, and the beam itself is comparable to the diameter acoustic emitters, in this case, when analyzing attenuation, processes refraction and wave divergence front can be neglected.

The attenuation coefficient, as defined by GOST, is a partial decrease in the amplitude of a plane wave per unit length of propagation in the medium with its one-time passage on

any specific frequency. How Typically, the attenuation coefficient is expressed in dB m

-1 by multiplying the partial attenuation by a factor of 8.686.

When specifying attenuation parameters

in liquids, data are given on their change with frequency. The attenuation coefficient at frequency f is defined as where f is expressed in MHz;

is the attenuation coefficient 1 MHz; b – constant determined by curve obtained by smoothing experimental data by method least squares. This option refers only to attenuation in the medium; it is not associated with reflection losses from interfaces and with decreasing signal due to diffraction. Sometimes underestimating the properties of scatterers can lead to erroneous results in attenuation estimates.

In IPTS based on a speed meter sound system IZZ-1, a prototype of the device was developed with simultaneous speed measurement sound and attenuation coefficient, description of the layout and algorithm of its operation given in article [32]. As noted above, the layout is equipped with additional acoustic channel with a piezoelectric receiver, which is located at an angle to incident beam, which allows simultaneous measurement of forward and backward scattering. In fact, we have a multifunctional acoustic device that allows us to simultaneously determine the speed of sound and coefficients attenuation, scattering and absorption.

Metrological support for similar instruments still requires additional research, and this task should be solved by joint efforts of developers of acoustic meters with the involvement of specialists VNIIFTRI.

Conclusions. Recommendations were made for application of GOST for acoustic measuring the speed of sound in a liquid,attenuation and scattering and states, that metrological characteristics newly created acoustic measuring instruments operating in in situ conditions are at the level of State primary standards and these standards need to be improved.

Metrological support for multifunctional acoustic means for measuring the speed of sound, attenuation, scattering and absorption, – The issue is complex and far from resolved, which requires additional research.